\documentclass[aps,pra,twocolumn,showkeys,nofootinbib,superscriptaddress]{revtex4-2}

\usepackage{amsmath,amssymb,bm}
\usepackage[hidelinks]{hyperref}
\usepackage{microtype}

\begin{document}

\title{Comment on Temperature change can solve the Deutsch--Jozsa problem: An exploration of thermodynamic query complexity}

\author{Ridha Horchani}
\email{horchani@squ.edu.om}
\affiliation{Department of Physics, College of Science, Sultan Qaboos University, Muscat, Oman}

\date{July 2, 2026}

\begin{abstract}
We show that the readout analysis in Phys. Rev. A \textbf{113}, 012420 (2026) does not establish the claimed resource accounting of one thermal query followed by many statistically useful probe samples. A CNOT fanout of a diagonal post-query probe broadcasts its one-qubit marginal but produces perfectly correlated registers rather than independent copies. Consequently, neither the trace distance nor the relative entropy between the balanced and constant hypotheses increases, and repeated measurements of the ancillas provide only one Bernoulli observation. Independent samples instead require repeated preparation and heat exchange, which are repeated queries under the definition adopted in the original paper. We also show that the sample lower bound stated there does not follow from Pinsker's inequality: the inequality gives the opposite ordering for the reciprocal relative entropy. The thermal-kickback mechanism may still encode the decision in the probe temperature, but the one-query readout claim and the quoted necessity of approximately 116 samples require correction.
\end{abstract}

\maketitle

In Ref.~\cite{Xuereb2026}, a Boolean function is encoded in the energetic structure of a thermal-machine oracle and a probe qubit is used to query that oracle through a heat exchange. The post-interaction probe population differs according to whether the function is balanced or constant. The paper further argues that the Deutsch--Jozsa decision can be implemented with one thermal query and a number of probe samples that is independent of the input size. Two points in the readout argument prevent that conclusion. First, the proposed CNOT construction does not generate independent samples from one queried probe. Second, the lower bound stated in Eq.~(7) of Ref.~\cite{Xuereb2026} does not follow from the Chernoff--Stein and Pinsker inequalities.

Let the two hypotheses of interest be denoted by $h\in\{B,C\}$, where $B$ is the balanced case and $C$ is the least distinguishable of the two constant cases. Since the post-query probe is diagonal in the energy basis, write
\begin{equation}
 \rho_h=p_h\lvert0\rangle\!\langle0\rvert+(1-p_h)\lvert1\rangle\!\langle1\rvert .
 \label{eq:probe}
\end{equation}
Ref.~\cite{Xuereb2026} proposes that a single copy of this state can be ``unitarily reconstruct[ed]'' in pure ancillas by CNOT gates, thereby allowing repeated measurements. Consider $M$ ancillas initially in $\lvert0\rangle^{\otimes M}$ and apply a CNOT from the probe to every ancilla. The exact output is
\begin{equation}
 \sigma_h^{(M)}=
 p_h\lvert0^{M+1}\rangle\!\langle0^{M+1}\rvert
 +(1-p_h)\lvert1^{M+1}\rangle\!\langle1^{M+1}\rvert .
 \label{eq:broadcast}
\end{equation}
Every individual register has the marginal state $\rho_h$, but the joint state is not $\rho_h^{\otimes(M+1)}$. The outcomes are perfectly correlated: an energy-basis measurement produces either $00\cdots0$ or $11\cdots1$. Thus the $M+1$ recorded bits constitute one Bernoulli draw copied $M$ times, not $M+1$ independent draws. In particular, their empirical frequency is either zero or one and does not converge to $p_h$ as $M$ grows.

This conclusion can be stated directly in hypothesis-testing language. For the trace distance,
\begin{align}
 T\!\left(\sigma_B^{(M)},\sigma_C^{(M)}\right)
 &\equiv \frac12\left\|\sigma_B^{(M)}-\sigma_C^{(M)}\right\|_1 \\
 &=\lvert p_B-p_C\rvert
 =T(\rho_B,\rho_C),
 \label{eq:trace}
\end{align}
and for the relative entropy,
\begin{align}
 D\!\left(\sigma_B^{(M)}\middle\|\sigma_C^{(M)}\right)
 &=p_B\ln\frac{p_B}{p_C}
 +(1-p_B)\ln\frac{1-p_B}{1-p_C} \\
 &=D(\rho_B\|\rho_C).
 \label{eq:relative}
\end{align}
The CNOT fanout is an isometric embedding, so it preserves rather than amplifies the distinguishability of the two input states. By contrast, $M$ genuinely independent probes would satisfy
\begin{equation}
 D\!\left(\rho_B^{\otimes M}\middle\|\rho_C^{\otimes M}\right)
 =M D(\rho_B\|\rho_C),
 \label{eq:iid}
\end{equation}
which is the extensivity required by an independent-sample Chernoff--Stein analysis. The distinction is not a no-cloning objection. Commuting states can be broadcast, and Eq.~\eqref{eq:broadcast} is a valid broadcast state \cite{Barnum1996}. The issue is that broadcasting equal marginals does not create statistically independent copies and does not create additional information about the unknown population.

The alternative readout discussed in Ref.~\cite{Xuereb2026} is to measure the probe, reset the probe and machine, and repeat the experiment. This procedure can generate independent data, but it changes the query count. The original paper defines a query as a heat exchange between the probe and the thermal-machine oracle. After a measurement and reset, producing another post-query state $\rho_h$ requires another such heat exchange. Therefore,
\begin{equation}
 \begin{split}
 M\ \text{independent post-query samples}\\
 \Longrightarrow\; M\ \text{thermal queries}.
 \end{split}
 \label{eq:counting}
\end{equation}
for the protocol actually specified. The two proposed readings consequently have complementary limitations: CNOT fanout uses one query but yields only one independent observation, whereas reset-and-repeat yields $M$ independent observations but uses $M$ queries. No protocol in Ref.~\cite{Xuereb2026} supplies $M$ independent samples after a single heat exchange.

There is also an independent problem with the sample bound. Ref.~\cite{Xuereb2026} states that the required sample number $n^{*}$ is lower bounded by
\begin{equation}
 n^{*}>\frac{\ln(1/\delta)}{D(\rho_B\|\rho_C)},
 \label{eq:starting}
\end{equation}
and then uses Pinsker's inequality together with the distinguishability condition $T(\rho_B,\rho_C)>t$. Even accepting Eq.~\eqref{eq:starting} as the starting point, Pinsker's inequality gives
\begin{equation}
 D(\rho_B\|\rho_C)\geq 2T(\rho_B,\rho_C)^2>2t^2,
 \label{eq:pinsker}
\end{equation}
so that
\begin{equation}
 \frac{\ln(1/\delta)}{D(\rho_B\|\rho_C)}
 <\frac{\ln(1/\delta)}{2t^2}.
 \label{eq:reciprocal}
\end{equation}
A lower bound on $n^{*}$ by the left-hand side of Eq.~\eqref{eq:reciprocal}, combined with an upper bound on that left-hand side, does not imply
\begin{equation}
 n^{*}>\frac{\ln(1/\delta)}{2t^2}.
\end{equation}
Hence Eq.~(7) of Ref.~\cite{Xuereb2026} is not a consequence of Pinsker's inequality, and the statement that $\delta=t=0.1$ requires at least $50\ln 10\simeq116$ samples is unsupported. At most, if one replaces the asymptotic Chernoff--Stein relation by an estimate proportional to $\ln(1/\delta)/D$, the Pinsker substitution supplies a conservative upper estimate for that expression, not a necessary lower bound. More generally, the Chernoff--Stein lemma concerns the asymptotic exponent of one hypothesis-testing error at fixed tolerance for the other; it is not, without additional finite-size analysis, the exact finite-sample inequality used in Ref.~\cite{Xuereb2026} \cite{CoverThomas,NielsenChuang}.

These observations do not invalidate the population-transfer calculation or the possibility that thermal kickback encodes the balanced and constant cases in different probe temperatures. They also do not exclude a query complexity that remains constant in the input size for a fixed target error, provided the population separation remains bounded away from zero. They do, however, alter the operational conclusion. One queried mixed probe gives one stochastic observation. Producing the independent observations required for temperature estimation entails repeated uses of the heat-exchange oracle, unless a different readout channel with a demonstrated information gain is supplied. Accordingly, the claims of one query followed by many useful samples, the comparison based on that resource separation, Eq.~(7), and the numerical ``at least 116 samples'' statement should be corrected.

\end{document}